
\input phyzzx.tex

\def\qgg {\Delta_{\hat g}}
\def\d {\partial}

\def\gg {\hat g}

\def\a {\alpha}

\def\t {\theta}

\def\b {\beta}
\def\G {\Gamma}
\def\F {\Phi}
\nopagenumbers

\font\tenbf=cmbx10
\font\tenrm=cmr10
\font\tenit=cmti10
\font\elevenbf=cmbx10 scaled\magstep 1
\font\elevenrm=cmr10 scaled\magstep 1
\font\elevenit=cmti10 scaled\magstep 1

\font\ninerm=cmr9

\line{\hfil }
\vglue 1cm
\hsize=6.0truein
\vsize=8.5truein
\parindent=3pc
\baselineskip=10pt
\centerline{\tenbf A CONTINUUM APPROACH TO 2D-QUANTUM GRAVITY FOR C$>$1}
\vglue 5pt
\centerline{\ninerm }
\vglue 1.0cm
\centerline{\tenrm M. MARTELLINI \foot{On leave of absence from
Dipartimento di Fisica, Universit\`a
di Milano, Milano, Italy and I.N.F.N., Sezione di Pavia, Italy}}
\baselineskip=13pt
\centerline{\tenit I.N.F.N., Sezione di Roma "La Sapienza", Roma, Italy}
\vglue 0.3cm
\centerline{\tenrm M. SPREAFICO}
\baselineskip=13pt
\centerline{\tenit Dipartimento di Fisica, Universit\`a di Milano, Milano,
Italy}
\vglue 0.3cm
\centerline{\tenrm and}
\centerline{\tenrm K. YOSHIDA}
\baselineskip=13pt
\centerline{\tenit Dipartimento di Fisica, Universit\`a di Roma, Roma, Italy
and
I. N. F. N., sezione di Roma, Italy}
\vglue 0.8cm
\centerline{\tenrm ABSTRACT}
\vglue 0.3cm
{\rightskip=3pc
 \leftskip=3pc
 \tenrm\baselineskip=12pt
 \noindent
A Lagrangian continuum model for a string theory with central charge
c$>$1 is formulated by incorporating Weyl and diffeomorphism gauge fixing.
In particular the tachyon scattering amplitudes are deduced generalizing the
standard c$\le$1 computation.
\vglue 0.8cm }
\line{\elevenbf 1. Lagrangian formulation \hfil}
\baselineskip=14pt
\elevenrm
\vglue 0.4cm

Despite the remarkable progress of last years in understanding the quantum
theory of gravity in 2 dimensional space-time, a serious problem still remains,
connected with the inability of going beyond the matter central charge at value
$c=1$.

A possible way to overcome this difficulty lies in a slight
generalization of the David, Distler and Kawai$^1$ (DDK)
model in the conformal gauge, obtained by
introducing a new scalar field coupled to the "Liouville" one.
This generalization can be motivated by an observation on the procedure
used to deduce the partition function for 2DQG by summing on smooth
2-dimensional random surfaces, from which one is induced to introduce
in the path integral a new Jacobian term to fix all the gauge freedom in
the right way. Unfortunately by this way we define a infinite set of model,
characterized by the value of a new free parameter.

In this note, after resuming the meaningful result of this approach, we
proceed to compute the tachyon correlation functions of the model.
In the simplest case the kinematical and regularity constraints
allow to gain more
information on the free parameter of the theory.

Taking care of both diffeomorphisms and Weyl symmetries in
quantizing the non critical string, one is led to a simple extension of
the DDK$^1$ ansatz where the total gauge fixing jacobian give rise to
the following combined Liouville-Weyl conformal field theory$^{2,3}$
$$
\eqalign{
&S_{\scriptscriptstyle{WL}} =
{1\over 8 \pi}
\int d^2\xi \sqrt{\gg} [-M_{a,b}\Phi^a\qgg\Phi^b-Q_a\Phi^a R_{\gg}]\cr
&M_{a,b}\equiv\left(\matrix{1&B\cr
B&-1\cr}\right)\cr}
\eqn\ea
$$

Here indices are summed when repeated and take values
$1$ and $2$, $\F^1$ is the Liouville field, $\F^2$ is the
bosonized Weyl-ghost field and $B$, $Q_1$ and $Q_2$ are general
free parameters.
In the following, we shall take the attitude to understand \ea\ as the
defining theory, no matter the way followed to get it.

Holomorphic energy-momentum tensor associated with \ea\ reads
locally as
$$
T = -{1\over 2}[M_{a,b}\d\Phi^a\d\Phi^b+Q_a(\d^2)\Phi^a]
$$

The corresponding central charge is
$$
\eqalign{
&c_{\scriptscriptstyle{W+L}}=2+3M^{a,b}Q_aQ_b\cr
&M^{a,c} M_{c,b}=\delta^a_b\cr}
$$
and Weyl invariance of the whole system requires
$$
c_{\scriptscriptstyle{W+L}}+c_{\scriptscriptstyle{M}}-26=0
$$
which gives
$$
\eqalign{
&Q_1^2 + 2B Q_1 Q_2 - Q_2^2 + {M\over 3}(c_{\scriptscriptstyle{M}} - 24) = 0\cr
&M = - det(M_{a,b}) = 1 + B^2\cr}
\eqn\eb
$$

Eq. \eb\ represents the conformal invariance of the whole
system as a relation between the central charges of the two fields
$\F^1$ and $\F^2$. One can choose the special class of solutions of \eb\
setting up the ansatz
$$
\eqalign{
&Q_2(B,c_{\scriptscriptstyle{M}})=
\sqrt{{1+c_{\scriptscriptstyle{M}}B}\over 3}\cr
&Q_1(B,c_{\scriptscriptstyle{M}})=
-{1\over \sqrt{3}} \left[ B\sqrt{1+Bc_{\scriptscriptstyle{M}}}-
\sqrt{1 + B^2}
\sqrt{25 + (B-1)c_{\scriptscriptstyle{M}}} \right]\cr}
$$

To gain more information on the behaviour of the model and to test
its validity we proceeded to calculate string susceptibility.$^2$
To do that we follow the method of DDK who used a constant rescaling argument
of the Liouville field, that is we evaluate the variation of the partition
function for a variation of the area of the Riemann surface. Our ansatz is
that this area is expressed by the usual relation: $A=\int d^2 x \sqrt{\hat
g} e^{\a_1\F^1}$, where the Weyl field does not appear.
This is because that if Weyl symmetry has been fixed then
that is equivalent to fix the area of the Riemann surface and so this area
can not vary any more, vice versa if the area varies then the Weyl symmetry
has not been fixed and so the path integral must remain ambiguous.

We can then calculate the string susceptibility by using
$$
\Gamma(h) = \chi(h) {Q_1\over 2\a_1} +2
\eqn\gamma
$$
where $\a_1$ is the momentum of the Liouville field, and its value can be
obtained by considering that the conformal dimension of the gravitational
dressing $e^{\a_1\F^1}$ for the area operator, where again the
Weyl field does not appear, must be one:
$$
\Delta_0 [e^{\alpha_1\phi_1 }]=-{1\over 2} \alpha_1 M^{1,b}(\alpha_b+Q_b)=1
\eqn\ec
$$
from which we get (we chose the solution $\a_1=\a_{1+}$ for the
reason understood in ref. 2)
$$
\eqalign{
\a_1 &= - {1\over 2} \left[Q_1+BQ_2-\sqrt{(Q_1+BQ_2)^2-8M}\right]\cr
&=- {1\over 2}\left[ \sqrt{M(Q_2^2+8-{c_{\scriptscriptstyle{M}}\over 3})}
-\sqrt{M(Q_2^2-{c_{\scriptscriptstyle{M}}\over 3})}\right]\cr}
\eqn\ee
$$

Inserting the above solutions the final
expressions for $\a_1$ and $\G$ are

$$
\eqalign{
&\a_1(B,c_{\scriptscriptstyle{M}})=
-{\sqrt{1 + B^2}\over 2 \sqrt{3}}
\left[\sqrt{25+(B-1)c_{\scriptscriptstyle{M}}}
-\sqrt{1+(B-1)c_{\scriptscriptstyle{M}}} \right]\cr
&\Gamma (B,c_{\scriptscriptstyle{M}},h)={2(1-h)\over \sqrt{1+B^2}}
{B\sqrt{1+Bc_{\scriptscriptstyle{M}}}-
\sqrt{1+B^2}\sqrt{25+(B-1)c_{\scriptscriptstyle{M}}}\over
\sqrt{25+(B-1)c_{\scriptscriptstyle{M}}}-
\sqrt{1+(B-1)c_{\scriptscriptstyle{M}}}}+2\cr}
\eqn\ed
$$

Studying $\Gamma$ as a function of $B$, we find$^2$
that it is a real quantity for
$B>1-{1\over c_{\scriptscriptstyle{M}}}$
if $c_{\scriptscriptstyle{M}}>0$.

To proceed to calculate the correlation functions of our model we couple it
to a conformal matter represented by the following action
$$
S_M={1\over 8\pi}
\int d^2\xi \sqrt{\gg} [-X^a \qgg X_a]
$$
where $X^a$ are $d$ bosonic field and the central charge of the matter system
is
$$
c_{\scriptscriptstyle{M}}=d
$$

Matter vertex operators $V_{k^a} = e^{ik_a X^a}$ will be dressed by
the Liouville field alone
$$
T_{k^a}=e^{i k_a X^a} e^{\b_1 \F^1}
$$
and again conformal invariance requires (we set $ k_a=k~ \forall a$)
$$
\eqalign{
&\Delta [T_{k^a}]=1\cr
&{1\over 2} k_a k^a - {1\over 2} \b_a M^{a,b}(\b_b+Q_b)=1\cr
&{d\over 2} k^2 - {1\over 2} \b_1 M^{1,b}(\b_b+Q_b)=1~~(\b_2=0)\cr
&\b^2+(Q_1+BQ_2)\b +M[2-d k^2)] =0~~(\b:=\b_1)\cr}
$$
with solution
$$
\eqalign{
&\b(B,k)=\cr
&=-{\sqrt{1+B^2}\over 2\sqrt{3}} \left[
\sqrt{25+d(B-1)}-\sqrt{1+d(B-1)+12 d k^2}
\right]\cr}
$$

\vglue 0.6cm
\line{\elevenbf 2. Tachyon scattering amplitudes \hfil}
\vglue 0.4cm
To compute tachyon scattering amplitudes we follow the procedure defined
firstly by Gupta et al.$^4$ and then stated in a more complete way by
Di Franceso and Kutasov.$^5$ This procedure is really introduced for a $c\leq
1$
Liouville theory, anyway
we can generalize it to our model and we get the result below
(where the vector index $a$ in $k^a_i$ has been omitted)
$$
\eqalign{
A(k_1,...,k_n)&=<T_{k_1},...,T_{k_n}>\cr
&={2 \sqrt{\pi}\over \a_1} A^{-1-s}
\int D\Phi' DX' e^{-S_{WL}[\Phi']-S_M[X']}
\left(\int d^2 z e^{\a_1\Phi'_1 (z)}\right)^s T_{k_1} ... T_{k_n}\cr}
\eqn\fcorr
$$
here $A$ is the area of the Riemann surface,
$\Phi'$ and $X'$ represent the non-zero modes of the fields
and $s$ is given by
$$
\sum_{i=1}^n \b(k_i)+s \a_1=-Q_1
\eqn\conds
$$

A second condition to take in account is given by the
charge neutrality of the matter theory; in the actual case,
where the background charge of the matter sector is
zero, this traduces simply in the following requirement
(that is there are not screening charges)
$$
\sum_{i=1}^n k_i = 0
\eqn\condk
$$

It is clear that we can compute \fcorr\ when $s$ is a positive integer since
the amplitude reduces to a free field one, in fact we obtain
$$
\eqalign{
A(k_1,...&,k_n)={2 \sqrt{\pi}\over \a_1} A^{-1-s}\cr
&\prod_{i=1}^n \prod_{j=1}^s \int d^2 t_j |x_i-t_j|^{-{2\over 1+B^2}\a_1\b_i}
\prod_{i'<i} |x_{i'}-x_i|^{-{2\over 1+B^2}\b_{i'}\b_i}
\prod_{j'<j} |t_{j'}-t_j|^{-{2\over 1+B^2}\a^2_1}\cr}
$$

Otherwise it becomes unclear how to perform this calculation
when $s$ is an arbitrary number. To overcome this difficulty
the idea of Kutasov and Di Francesco is based on the theorem of Carlson
which states that it
is possible,  under favorable conditions,
to continue  a function calculated for
an infinite set of positive integer values to the whole
complex positive half-plane.

Following this idea we start by computing \fcorr\ for $n=3$ and $s$ integer,
we obtain
$$
\eqalign{
A(k_1,k_2,k_3)&=<T_{k_1},T_{k_2},T_{k_3}>\cr
&={2\sqrt{\pi}\over \a_1}A^{-s-1}
\prod_{i=1}^s \int_C d^2 z_i |z_i|^{2a} |1-z_i|^{2b}
\prod_{j<i} |z_j-z_i|^{4c}\cr}
\eqn\pippo
$$
where
$$
\eqalign{
&a:=-{1\over 1+B^2}\a_1\b(k_1),~~~~b:=-{1\over 1+B^2}\a_1\b(k_2)\cr
&c:=-{1\over 2(1+B^2)} \a_1^2\cr}
$$

This integral presents a deep analogy with the famous Selberg integral,
which is also computed by using the Carlson theorem, and in fact the result,
indeed obtained by different method,$^6$ show very suggestive similarities.
It is clear that this result must underlie some conditions on the
three parameters $a,b,c$ to ensure the convergence of the integral.
We examine these conditions in the simple cases of $s=1$
and $s=2$.
We set:
$$
I(a,b,c,s)=\prod_{i=1}^s \int_C d^2 z_i |z_i|^{2a} |1-z_i|^{2b}
\prod_{j<i} |z_j-z_i|^{4c}
$$

We can evaluate directly (appendix B) the case $s=1$
$$
I(a,b,c,1)=\int_C d^2 z |z|^{2a} |1-z|^{2b}=\pi {\G(-a-b-1)\over \G(-a)\G(-b)}
{\G(a+1)\G(b+1)\over\G(a+b+2)}
$$
under the conditions
$$
Re~a>-1,~~Re~b>-1,~~Re(a+b)<-1
\eqn\conda
$$

For $s=2$
$$
I(a,b,c,2)=\int_C d^2 z_1 d^2 z_2 |z_1|^{2a} |z_2|^{2a}
|1-z_1|^{2b} |1-z_2|^{2b} |z_1-z_2|^{4c}
$$
even if a direct evaluation is not still known, we can deduce
the following convergence conditions (appendix A)
$$
\eqalign{
&Re~a>-1,~~Re~b>-1\cr
&-Min\left[{1\over 2}, Re(a+1), Re (b+1)\right]<Re~c<-Max\left[{1\over 2}
Re(a+b+1),Re(a+b+1)\right]\cr}
\eqn\condb
$$

It appears that the case of $s=1$ is fundamentally different from the
general case $s>1$, so we shall consider condition \condb\ rather than
condition \conda\ as representing the generic case (in the sense that
from it we can guess the conditions for the generic case).

\vglue 0.6cm
\line{\elevenbf 3. Explicit calculation: d=4 target space time case \hfil}
\vglue 0.4cm
So concentrating on the convergence conditions in the simplest case \condb,
we can look for a range for the free parameters which would ensure
it. In the actual case just three of parameters $B, d, k_1, k_2, k_3$ are still
free, $k_2$ being fixed by condition \condk\ ($k_2=-k_1-k_3$), and
$B, d, k_1, k_3$ related by \conds. Moreover  we must remember the condition
$B>1-{1\over c_{\scriptscriptstyle{M}}}$ which ensures the reality of $\G$.
Since we are interested in the case of a physical matter, for simplicity
we set $c=d=4$ (as the dimensions for the space time).
We then get the following expression for the quantity of interest:
$$
\eqalign{
\a_1(B)&=-{\sqrt{1 + B^2}\over 2 \sqrt{3}}  \left(\sqrt{21+4B}
-\sqrt{4B-3} \right)\cr
Q_1(B)&=-{1\over \sqrt{3}} \left[ B\sqrt{1+4B}- \sqrt{(1 + B^2)
(21 + 4B)} \right]\cr
\Gamma (B,h)&={2(1-h)\over \sqrt{1+B^2}}
{B\sqrt{1+4B}-\sqrt{(1+B^2)(21+4B)}\over
\sqrt{21+4B}-\sqrt{4B-3}}+2\cr
\b(B,k)&=-{\sqrt{1+B^2}\over 2\sqrt{3}} \left[
\sqrt{21+4B)}-\sqrt{4B-3+48 k^2}
\right]\cr}
$$
and
$$
\eqalign{
&a=f(B,k_1)\cr
&b=f(B,k_3)\cr
&f(B,k)=-{1\over 1+B^2}\a_1(B)\b(B,k)\cr
&=-{1\over 12}\left( \sqrt{21+4B}-\sqrt{4B-3}\right)
\left(\sqrt{21+4B}-\sqrt{4B-3+48k^2}\right)\cr
&c=-{1\over 2(1+B^2)} [\a_1(B)]^2\cr
&=-{1\over 24}\left(\sqrt{21+4B}-\sqrt{4B-3}\right)^2\cr}
$$

For the from eq. \conds\ for $s=2$ we get
$$
2\a_1+Q_1+\b(k_1)+\b(k_3)+\b(-k_1-k_3)=0
$$
which, after the substitution of above quantities, becomes
$$
2\left(B\sqrt{{1+4B\over 1+B^2}}+3\sqrt{21+4B}\right)
-f(B,k_1)-f(B,k_3)-f(B,-k_1-k_3)=0
$$

Analytic solution of this equation is rather hard, and even harder is the
solution of the set of disequations following from condition \condb,
the only way to handle these constraints being by numerical evaluation.
On this way we immediately find that to satisfy the necessary condition
$c>-{1\over 2}$, $B$ must be greater than ${3\over 2}$. We then start by the
minimum value for $B$ getting the correspondent value for $c$:
$$
\eqalign{
&B=B_0={3\over 2}+\epsilon=1.50001\cr
&c=-0.499998\cr}
$$

Keeping fixed the value of $B=B_0$, we can now find a
minimum value also for the momenta $k_i$. In fact from
\condb\ we see that $-a-1$ ($-b-1$) must be less than $c$, so evaluating the
variation of $-a-1$ ($-b-1$) in function of $k_1$ ($k_3$) for $B=B_0$
we find that $k_i>k_{\scriptscriptstyle{min}}=0.433010$.
Then fixed $k_1=k_{\scriptscriptstyle{min}}+\epsilon=0.433011$
we can look for a value for $k_3$ by solving numerically \conds\ , we get
$k_3\sim 0.690571$, which satisfy the condition stated above for $k_i$. At
this point we analyze the complete condition \condb\ for these values,
and we find that the condition is verified and the final values of the
parameters are:
$$
\eqalign{
&a=-0.500001\cr
&b=-0.0311387\cr
&c=-0.499998\cr
&\G=0.270978\cr}
$$

We can now try to vary the value of $k_1$ (and then the one of $k_3$) by
keeping
$B=B_0$ fixed, to define the whole range for the momenta. The stronger
condition
we must keep in mind is now $c<-a-b-1$. What we find is that increasing
the value of $k_1$ the one of $k_3$ decreases, so we can do it until $k_3$
reach the minimum value allowed for the momenta, $k_{\scriptscriptstyle{min}}$.
We find that (as was readily obvious from the symmetry of condition \conds\
when $k_2=-k_1-k_3$) the maximum value for the momenta is
$k_{\scriptscriptstyle{max}}\sim 0.690571$. We note also that the condition
$c<-a-b-1$ is always satisfied by the couple of values for ($k_1, k_3$) in the
range $[k_{\scriptscriptstyle{min}},k_{\scriptscriptstyle{max}}]\times
[k_{\scriptscriptstyle{min}},k_{\scriptscriptstyle{max}}]$, so it does not
introduce more constraints. More for all these couples of values
the whole condition \condb\ is satisfied, and so they represents a good set
of solutions of our problem. Here below some sets of allowed values.
$$
\eqalign{
&B=B_0=1.50001\cr
&c=-0.499998\cr}
$$
$$
\matrix{
k_1&k_3&a&b&|k_1-k_3|\cr
0.433011&0.690571&-0.500001&-0.0311387&0.25756\cr
0.5&0.625464&-0.381965&-0.152847&0.125464\cr
0.563012&0.563012&-0.267956&-0.267956&0\cr}
$$

We can now try to vary the value of $B$ from the value $B_0$, looking for new
sets of solutions. We immediately see that the value of $c$ increases by
increasing $B$, with the limit $\lim_{B\to\infty} c =0^-$, so the first
condition $c>-{1\over 2}$ is always satisfied. By fixing successively
different increasing values of $B=B_n, n=1,2,...$
we look for the minimum values for the
$k_i$ by considering again condition $-a-1<c$ ($-b-1<c$). We find that
$k_{\scriptscriptstyle{min},n}$ decreases respect to the above value set for
$B=B_0$. Then fixing $k_1=k_{\scriptscriptstyle{min},n}$ the
correspondent value for $k_3$, as deduced from \conds, increases.
We finally must consider the last condition $c<-a-b-1$, differently from the
case of $B=B_0$, now this condition must be kept in account. In fact while
for the
case of $B=B_0$, all the allowed values of ($k_1,k_3$) satisfied this
condition,
now we find that as the value of $c$ increases not all the couples
($k_1,k_3$) give values of $a$ and $b$ that satisfy it: the range
for the momenta reduces increasing the value of $B$.
What happens increasing $B$ is that
once one have fixed an acceptable value for $k_1$
then the value deduced using eq. \conds\
for $k_3$ is often too big to satisfy the condition $c<-1-a-b$.

This traduces in the result that a maximum value for $B$ exists.
More over, by numerical evaluation we find that the maximum value for $B$
is really close to the minimum one, in fact $B=1.6$ is still acceptable while
$B=1.7$ is not, so the set of allowed values for $B$ is now almost reduced
to a point as we need. To summarize for $B=1.6$ we have, for instance:
$$
\eqalign{
&B=B_1=1.6\cr
&c=-0.479005\cr}
$$
$$
\matrix{
k_1&k_3&a&b&|k_1-k_3|\cr
0.53&0.594627&-0.318033&-0.203711&0.064627\cr
0.55&0.574767&-0.282916&-0.239095&0.024767\cr
0.562395&0.562395&-0.261029&-0.261029&0\cr}
$$.

\vglue 0.6cm
\line{\elevenbf 4. Acknowledgements \hfil}
\vglue 0.4cm
K. Yoshida wishes to thank for stimulating discussions H. Kawai and
K. Fujikawa, M. Spreafico is grateful to Vl. S. Dotsenko and E. Montaldi
for helpful conversations, and M. Martellini acknowledges partial support
from MPI-40\%.
\vglue 0.6cm
\line{\elevenbf 5. References \hfil}
\vglue 0.4cm
\item{1.} J. Distler and H. Kawai, {\elevenit Nucl. Phys.}
{\elevenbf B321} (1989) 509.
\item{2.} M. Martellini, M. Spreafico and K. Yoshida,
{\elevenit Mod. Phys. Lett.} {\elevenbf A7} (1992) 1281.
\item{3.} J. Cohn and V. Periwal,
{\elevenit Phys. Lett.} {\elevenbf B270} (1991) 18.
\item{4.}
A. Gupta, S. P. Trevedi and M. B. Wise, {\elevenit Nucl. Phys.}
{\elevenbf B340} (1990) 475.
\item{5.} P. Di Francesco and D. Kutasov, {\elevenit Phys. Lett.}
{\elevenbf B261} (1991) 385.
\item{6.} Vl. S. Dotsenko and V. A. Fateev, {\elevenit Nucl. Phys.}
{\elevenbf B240} (1984) 312.
\vfill
\vglue 0.6cm
\line{\elevenbf 6. Appendix A \hfil}
\vglue 0.4cm
We consider the integral:
$$
I[a,b,c,s]:= \prod_{i=i}^s \int_{C_{i}} d^2 z_i |z_i|^{2a} |1-z_i|^{2b}
\prod_{j<i} |z_j-z_i|^{4c}
\eqn\uno
$$
that can be written in the following alternative ways:

cartesian coordinates
$$
I[a,b,c,s]:= \prod_{i=i}^s \int_{-\infty}^{+\infty} dx_i dy_i
(x_i^2+y_i^2)^a [(1-x_i)^2+y_i^2]^b \prod_{j<i} [(x_j-x_i)^2+(y_j-y_i)^2]^{2c}
$$

polar coordinates
$$
I[a,b,c,s]:= \prod_{i=i}^s \int_0^{2\pi} d\t_i \int_0^{+\infty} dr_i r_i^{2a+1}
(1+r^2_i+2r_i cos\t_i)^b \prod_{j<i} [r_j^2+r_i^2+2r_j r_i cos(\t_j-\t_i)]^{2c}
$$

Convergence of integral \uno\ depends on the behaviour of the integrand in
function of the parameters $a, b, c, s$ in the singular points, which can be
identified in $z_i=0\ne z_k~ \forall k\ne i;~ z_i=1\ne z_k~ \forall k\ne i;~
z_i=+\infty\ne z_k~ \forall k\ne i.~$ Then also the cases when one or more
different variables assume the same value, as for example:
$z_i=z_k\ne 0,1,\infty~ \forall k;~ z_i=0~ \forall i;~ z_i=1~ \forall i;~
z_i=\infty~ \forall i$.

We start by simplicity to consider the case of $s=1$ and $s=2$.

For $s=1$ integral \uno\ becomes
$$
\eqalign{
I[a,b,c,1]&=  \int_{C} d^2 z |z|^{2a} |1-z|^{2b}\cr
&= \int_{-\infty}^{+\infty} dx~ dy(x^2+y^2)^a [(1-x)^2+y^2]^b\cr}
$$
and can be calculated (appendix B) under the conditions:
$$
Re~a>-1,~Re~b>-1,~Re(a+b)<-1
$$

We can obtain the same conditions by considering the behaviour in
$z=0$ and $z=1$
$$
z=0 ~~~I\sim z^{2a+2}
$$
where the second term ($2$) in the exponent comes from the fact that we
integrate over two variables. From this behaviour we get the condition
$$
Re~a>-1
\eqn\ca
$$

Then by using the symmetry on the variable change $z\to 1-z$,
we can evaluate the behaviour in $z=1$ getting the condition for $b$,
$$
Re~b>-1
\eqn\cb
$$

In $z=\infty$ we have
$$
z=\infty, ~~~I\sim z^{2a+2b+2}
$$
from which we get the last condition $Re(a+b)<-1$.

For $s=2$ integral \uno\ becomes
$$
\eqalign{
I[a,b,c,2]&=  \int_{C} d^2 z_1 d^2 z_2 |z_1|^{2a} |z_2|^{2a}
|1-z_1|^{2b}
|1-z_2|^{2b} |z_1-z_2|^{4c}\cr
&= \int_{-\infty}^{+\infty} dx_1 dy_1 dx_1 dy_1 (x_1^2+y_1^2)^a
(x_2^2+y_2^2)^a
[(1-x_1)^2+y_1^2]^b  [(1-x_2)^2+y_2^2]^b\cr
&~~~~~~ [(x_1-x_2)^2+(y_1-y_2)^2]^{2c}\cr}
$$

Let proceed to evaluate the behaviour in the singular points. If just one of
the variable is zero, integral behaviour is the same as in the case above
from which the condition
$Re~a>-1$, in the same way for $z_i =1$ we can deduce the condition
$Re~b>-1$.
If just one of the variable goes to infinity we have
$$
z_1 =\infty,~z_2=k,~~~I\sim (z_1 z_2)^{2a+2b+4c+2}
$$
(from now on $k$ represents a finite constant different from zero and one)
from which the condition
$$
Re~c<-{1\over 2}Re(a+b+1)
\eqn\cc
$$

Let now evaluate what happens when both the variables have singular values.
If both the variables are zero, integral behaviour is the following
$$
z_1=0,~z_2=0~~~I\sim (z_1 z_2)^{2a+2a+4c+2+2}
$$
from which we deduce the condition
$$
-Re(a+1)<Re~c
\eqn\cd
$$
and as above, for $z_1=z_2=1$ we get the condition
$$
-Re(b+1)<Re~c
\eqn\ce
$$

If both the variables go to infinity, we have
$$
z_1=z_2=\infty,~~~I\sim (z_1 z_2)^{2a+2a+2b+2b+4c+4}
$$
from which the condition
$$
Re~c<-Re(a+b+1).
$$

If, finally, both the variables have
the same not singular value,
we get
$$
z_1=z_2=k,~~~I\sim (z_1- z_2)^{4c+2}
$$
in this case the term $2$ comes from the fact that singularities
occurs from values not of the variables but of their differences, that is
there are two, not four degrees of freedom. From this we get the following
condition
$$
Re~c>-{1\over 2}
$$

We can resume all the conditions as follows
$$
-Min\left[{1\over 2},Re(a+1),Re(b+1)\right]<Re~c<-Max \left[
{1\over 2}Re(a+b+1), Re(a+b+1)\right].
$$

\vglue 0.6cm
\line{\elevenbf 7. Appendix B \hfil}
\vglue 0.4cm
We can calculate directly the integral in the case $s=1$,
$$
\eqalign{
I[a,b,c,1]&=  \int_{C} d^2 z |z|^{2a} |1-z|^{2b}\cr
&= \int_{-\infty}^{+\infty} dx~ dy(x^2+y^2)^a [(1-x)^2+y^2]^b\cr}
$$

We start by rewriting the integral by using the following propriety of
function $\G(x)$
$$
\G(x)=s^x\int_0^{+\infty} dt~e^{-sx}t^{x-1}~~~~Re~s>0,~~Re~t>0
$$
we get under the conditions: $~Re~a<0,~~Re~b<0$,
$$
\eqalign{
&={1\over \G(-a) \G(-b)} \int_0^{+\infty} du~dv~ u^{-a-1} v^{-b-1}
\int_{-\infty}^{+\infty}
dx~dy~ e^{-u(x^2+y^2)} e^{-v[(x-1)^2+y^2]}\cr
&={1\over \G(-a) \G(-b)} \int_0^{+\infty} du~dv~u^{-a-1} v^{-b-1}e^{-v}
\int_{-\infty}^{+\infty}
dx~dy~ e^{-(u+v)x^2+2xv} e^{-(u+v)y^2}\cr
&={1\over \G(-a) \G(-b)} \int_0^{+\infty} du~dv~u^{-a-1} v^{-b-1}e^{-v}
\sqrt{{\pi\over u+v}}e^{{v^2\over u+v}}\sqrt{{\pi\over u+v}}\cr
&={\pi\over \G(-a) \G(-b)} \int_0^{+\infty} du~dv~u^{-a-1} v^{-b-1}
{1\over u+v}e^{-{uv\over u+v}}\cr}
$$
if now we consider the variable change $u=tv,~~du=v~dt$
$$
\eqalign{
&={\pi\over \G(-a) \G(-b)} \int_0^{+\infty} dt~dv~v(tv)^{-a-1} v^{-b-1}
{1\over tv+v}e^{-{tv\over tv+v}}\cr
&={\pi\over \G(-a) \G(-b)} \int_0^{+\infty} dt~t^{-a-1} {1\over 1+t}
\int_0^{+\infty} dv~ v^{-a-b-1} e^{-{t\over t+1}v}\cr}
$$
using again the above propriety of $\G(x)$ function
we get, under the condition $~Re(a+b)<-1$,
$$
\eqalign{
&={\pi\over \G(-a) \G(-b)} \int_0^{+\infty} dt~ t^{-a-1} {1\over 1+t}
\G(-a-b-1)\left( {t\over 1+t}\right)^{a+b+1}\cr
&=\pi{\G(-a-b-1)\over \G(-a) \G(-b)} \int_0^{+\infty} dt~t^b (t+1)^{-a-b-2}\cr}
$$
here we can use the definition of function $B(x,y)$ and its expression in
function of $\Gamma(x)$
$$
B(x,y):=\int_0^{+\infty} dt~t^{x-1}(1+t)^{-x-y}={\G(x)\G(y)\over \G(x+y)},
{}~~~~ Re~x>0,~~Re~y>0
$$
and we get, under the conditions $~Re~a>-1,~~Re~b>-1$, the final result
$$
I(a,b,c,1)=\pi{\G(-a-b-1)\over \G(-a) \G(-b)} {\G(a+1) \G(b+1)\over \G(a+b+2)}
$$
that holds under the conditions
$$
Re~a>-1,~~Re~b>-1,~~Re(a+b)<-1
$$
which obviously comprehend also the conditions $a<0$ and $b<0$ since
$$
a+b<-1,~~~a>-1~~\to~~b<-1-a<-1+1=0.
$$
\vfill

\eject
\bye